\documentclass[a4paper,nofootinbib]{revtex4}
\usepackage{amsmath}
\usepackage{amssymb}
\usepackage{vmargin}
\setpapersize{A4}
\setmarginsrb{20mm}{25mm}{20mm}{25mm}{15pt}{15pt}{15pt}{30pt}

\newcommand{\vct}[1]{\ensuremath{\mathbf{#1}}}

\begin{document}

\author{Emmanuel Ranz}
\affiliation{Lyc\'ee Louis Barthou, 2 rue Louis Barthou, 64000 Pau, France}
\email[E-mail: ]{emmanuel.ranz@ac-bordeaux.fr}
\title{Magnetic field with cylindrical symmetry}

\date{\today}

\begin{abstract}
This short article contains the  derivation of a general formula for the magnetostatic field in the neighbourhood of an axis of symmetry.
 \end{abstract}

\maketitle

\section{Introduction}

There is a remarkable result in magnetostatic when the field posses cylindrical symmetry, namely, for points closed to the axis of symmetry the magnetic field is determined by the distance from the axis and the values it takes on this axis.
Demonstration of this result appears in many textbooks~\cite{Jackson} and is often left as an exercise; it consists in giving for each component, the first terms of an ascending series in powers of the distance from the axis.

Whereas the first terms are easily found by successive approximations method, the difficulty rapidly increases when we try to derive higher order terms by the same method.
The purpose of the following lines is to overcome this obstacle by a direct derivation of the general terms of each ascending series.

\section{Derivation}

We shall use the cylindrical coordinates $(r,\theta,z)$ with the $z$-axis for the symmetry axis.
The magnetic field, $\vct{B}$, has two components $B_r$ and $B_\theta$ which depend on $r$ and $z$.
It is searched in regions containing the axis of symmetry and where current are absent. 
With this former condition, Maxwell's equations for the magnetic field reduce to 

\begin{align}
  \label{eq:1}
  \vct\nabla\wedge\vct{B} &= 0,
  \\
  \label{eq:2}
  \vct\nabla\cdot \vct{B} &= 0.
\end{align}

The first equation allows the magnetic field to be derived from a scalar potential $\phi$ defined by

\begin{equation}
  \label{eq:3}
  \vct{B}=\vct\nabla\phi.
\end{equation}

The substitution of Eq.~\ref{eq:3} in Eq.~\ref{eq:2} leads to Laplace's equation for the potential
\begin{equation}
  \label{eq:4}
  \Delta\phi = 0.
\end{equation}

Note that the electrostatic field obeys the same equations and the final results for the magnetic field will thus be valid for the electrostatic field.

In cylindrical coordinates the Laplace's equation becomes

\begin{equation}
  \label{eq:5}
  \frac{1}{r}\frac{\partial}{\partial r}\left(\frac{r\partial\phi}{\partial r}\right)+\frac{\partial^2\phi}{\partial z^2}=0.
\end{equation}

The separation of variables by means of the product functions,

\begin{equation}
  \label{eq:6}
  \phi(r,z)=f(r)g(z),
\end{equation}

leads to the following system of differential equations where $k$ is the separation parameter

\begin{align}
  \label{eq:7}
  g'' + k^2 g &=0,\\
  \label{eq:8}
  r^2f''+rf'-r^2k^2f &=0.
\end{align}

The integration of the first equation gives $e^{\pm ikz}$. 
Solutions of the second equation are Bessel's functions of the first and second kind  with zero order~\cite{Abramowitz}, and whose argument, $ikr$, is purely imaginary~\footnote{Theses functions are called modified Bessel functions.}.
The Bessel's function of the second kind must be rejected because of its singularity at $r=0$. We thus retain only $J_0$, the Bessel's function of the first kind
\begin{equation}
  \label{eq:9}
  f(r)=J_0(ikr).
\end{equation}

Now we have the following set of elementary solutions of Eq.~(\ref{eq:5})

\begin{equation}
  \label{eq:10}
  J_0(ikr)e^{ikz},
\end{equation}

on which we shall expand the potential $\phi(r,z)$.

This potential is assumed to be a definite function on the $z$-axis, we write therefore

\begin{equation}
  \label{eq:11}
  \phi(0,z)=\varphi(z).
\end{equation}

Let's call $\hat\varphi(k)$, the Fourier transform of $\varphi(z)$:

\begin{equation}
  \label{eq:12}
  \varphi(z)=\int_{-\infty}^{+\infty}\hat\varphi(k)e^{ikz}dk.
\end{equation}

Since we have $J_0(0)=1$, the potential

\begin{equation}
  \label{eq:13}
  \phi(r,z) =\int_{-\infty}^{+\infty}\hat\varphi(k)J_0(ikr)e^{ikz}dk,
\end{equation}

obviously coincides with $\varphi$ on the $z$-axis and satisfies Laplace's equation.

This solution is with no doubt unique. 
Indeed, if we consider the exterior Dirichlet's problem in which the potential is known and equals to $\varphi$ on the surface of an infinite cylinder surrounding the $z$ axis, the unique solution of this problem would tend to our solution when we let the radius of the cylinder tend to zero.

Along the axis of symmetry, the the magnetic field has the only component $B_z$ which is supposed to be known. We thus write

\begin{equation}
  \label{eq:14}
  B_z(0,z)=b(z).
\end{equation}

The Fourier transform $\hat b(k)$ of $b(z)$ is related to $\hat\varphi(k)$ by

\begin{equation}
  \label{eq:15}
  \hat{b}(k)=ik\hat\varphi(k).
\end{equation}

Making use of the expansion of Eq.~(\ref{eq:13}), we derive the component $B_z$

\begin{equation}
  \label{eq:16}
  B_z
  =\frac{\partial\phi}{\partial z}
  =\int_{-\infty}^{+\infty}J_0(ikr)ik\hat\varphi(k)e^{ikz}dk
  =\int_{-\infty}^{+\infty}J_0(ikr)\hat{b(k)}e^{ikz}dk.
\end{equation}

Now, we must transform this formula into an expansion containing terms with $b(z)$ and its successive derivatives. 
We have, for this purpose, the following ascending series of $J_0$~\cite{Abramowitz}

\begin{equation}
  \label{eq:17}
  J_0(x)=\sum_{n=0}^{\infty}\frac{(-\frac14x^2)^n}{n!^2},
\end{equation}

which we report in Eq.~(\ref{eq:16}) after substituting $ikr$ for $x$; this yields to

\begin{equation}
  \label{eq:18}
  B_z=\sum_{n=0}^{\infty}\frac{(-\frac14r^2)^n}{n!^2}
  \int_{-\infty}^{+\infty}(ik)^{2n}\hat{b(k)}e^{ikz}dk.
\end{equation}

The integral is nothing else than the Fourier transform of $b^{(2n)}(z)$, we thus obtain the desired expansion for $B_z$:

\begin{equation}
  \label{eq:19}
  B_z=\sum_{n=0}^{\infty}\frac{(-\frac14r^2)^n}{n!^2}b^{(2n)}(z).  
\end{equation}

The $B_r$ component can be derived quite similarly, 

\begin{equation}
  \label{eq:20}
  B_r=\frac{\partial\phi}{\partial r}
  =\int_{-\infty}^{+\infty}ikJ'_0(ikr)\hat\varphi(k)e^{ikz}dk.
\end{equation}

Since we have  $J'_0=-J_1$ and the following ascending series for $J_1$

\begin{equation}
  \label{eq:21}
  J_1(x)=\frac{x}{2}\sum_{n=0}^{\infty}\frac{(-\frac14x^2)^n}{n!(n+1)!},
\end{equation}

we obtain

\begin{equation}
  \label{eq:22}
  B_z
  =-\frac{r}{2}\sum_{n=0}^{\infty}\frac{(-\frac14r^2)^n}{n!(n+1)!}b^{(2n+1)}(z).
\end{equation}

In particular, the first terms of the expansions of $B_r$ and $B_\theta$ components are easily found~\cite{Jackson}:

\begin{align}
  \label{eq:23}
  B_z(r,z)&=
  B_z(0,z)-\left(\frac{r^2}{4}\right)\frac{\partial^2 B_z(0,z)}{\partial z^2}\\
  B_r(r,z)&=
  -\left(\frac{r}{2}\right)\frac{\partial B_z(0,z)}{\partial z}
  +\left(\frac{r^3}{16}\right)\frac{\partial^3 B_z(0,z)}{\partial z^3}.
\end{align}

\end{document}